\documentclass[reprint,
superscriptaddress,
preprintnumbers,
 amsmath,amssymb,
pra,
floatfix,
]{revtex4-2}%4-1 for AIP

\usepackage{graphicx}% Include figure files
\usepackage{dcolumn}% Align table columns on decimal point
\usepackage{subfigure}
\usepackage{bm}% bold math
\usepackage[utf8]{inputenc}  % makes Å Ä Ö work properly.
\usepackage[unicode=true, bookmarks=false, breaklinks=false, pdfborder={0 0 1},colorlinks=false]{hyperref}

\usepackage{url}
\usepackage{natbib}
\usepackage{multirow}
\usepackage{float}
\usepackage{amsmath}

\usepackage{xcolor}
\usepackage{soul}
\usepackage{etoolbox}
\usepackage{derivative}
\usepackage{textcomp}

\begin{document}

\preprint{S.Ghorai \textit{et al.}/Instrument}

\title{A setup for direct measurement of the adiabatic temperature change in magnetocaloric materials}

\newcommand{\uppsala}{Department of Materials Science and Engineering,Uppsala University, Box 35, SE-751 03,Uppsala, Sweden}

\author{Sagar Ghorai}
\email[e-mail: ]{sagar.ghorai@angstrom.uu.se}
\affiliation {\uppsala}

\author{Daniel Hedlund}
\affiliation{\uppsala}

\author{Martin Kapuscinski}
\affiliation{\uppsala}

\author{Peter Svedlindh}
\affiliation{\uppsala}

\date{\today}

\begin{abstract}
In order to find a highly efficient, environment-friendly  magnetic refrigerant, direct measurements of the adiabatic temperature change $\Delta T_{adb}$ is required. Here, in this work a simple setup for the $\Delta T_{adb}$ measurement is presented. Using a permanent magnet Halbach array with a maximum magnetic field of $1.8$ T and a rate of magnetic field change of $5$ T/s, accurate determination of $\Delta T_{adb}$ is possible in this system. The operating  temperature range of the system is from $100$ K to $400$ K, designed for the characterization of materials with potential for room temperature magnetic refrigeration applications. Using the setup, the $\Delta T_{adb}$ of a first-order and a second-order compound have been studied. Results from the direct measurement for the first-order compound have been compared with $\Delta T_{adb}$ calculated from the temperature and magnetic field dependent specific heat data. By comparing results from direct and indirect measurements, it is concluded that for a reliable characterization of the magnetocaloric effect, direct measurement of $\Delta T_{adb}$ should be adopted.

\end{abstract}

\maketitle

%\tableofcontents

\section{Introduction}

Since 1997, after the observation of the giant magnetocaloric effect near room temperature in Gd$_5$(Si$_2$Ge$_2$),\cite{pecharsky1997giant} several thousands of research papers have been published in search of a suitable magnetic refrigerant which can be applied in room temperature magnetic refrigerators. Unfortunately, till today no  material has been found that is both commercially viable and environmentally friendly. The magnetocaloric effect (MCE) is an intrinsic property of a material originating  from the spin-phonon interaction of the material. In case of adiabatic conditions the total entropy of a system, being the sum of its phonon and spin (or magnetic) entropies, is conserved. Therefore, changing the magnetic entropy by the application or removal of an  magnetic field will change the phonon entropy and the  temperature of the system. This change of temperature is known as the adiabatic temperature change ($\Delta T_{adb}$) and the measurement of $\Delta T_{adb}$ is known as the direct measurement of the MCE. Isothermal magnetization measurements yielding information about the isothermal entropy change ($\Delta S_M$) upon application or removal of a magnetic field can also be used to estimate $\Delta T_{adb}$ if combined with temperature and magnetic field dependent heat capacity measurements ($C_H$). The measurement of $\Delta S_M$ and $C_H$ is known as the indirect measurement of the MCE. Owing to the widespread  availability of systems used for magnetization and specific heat measurements often indirect measurements of the MCE have  been reported. A keyword search on $19$th October $2022$, in the ``Web of Science" yields around $8929$ publications where the keyword ``magnetocaloric" is mentioned and among them only around $179$ publications mention ``magnetocaloric" and``direct" ``adiabatic temperature". These numbers are approximate but  give evidence of that reports of direct measurements of $\Delta T_{adb}$ are scarce. Moreover, several publications \cite{khovaylo2008adiabatic,moya2007cooling,sharma2007magnetocaloric,gschneidner2005recent,fujieda2004direct} report that $\Delta T_{adb}$ estimated from indirect measurements differs largely from that of  direct measurements owing to non-adiabatic conditions and approximations involved in indirect measurements. Pecharsky \textit{et al.}\cite{pecharsky1999magnetocaloric} have estimated the error involved in the indirect measurement process and found that it can be as large as $\sim15\%$ for elemental Gd. Furthermore, the values are highly sensitive to any approximation used in the calculation. By comparing direct and indirect MCE measurements, Pecharsky \textit{et al.} have also concluded that, in the indirect measurement, there will be around $1$ K to $1.5$ K uncertainty in the determination of $\Delta T_{adb}$ near room temperature\cite{pecharsky1996comparison}. It can therefore be argued that reliable estimattion of the $\Delta T_{adb}$ require direct measurements.

Here, in this work we  demonstrate a simple setup for the direct measurement of $\Delta T_{adb}$. Using this setup, we have compared the direct and indirect measurements of the MCE of the first-order material La$_{0.7}$Ca$_{0.3}$MnO$_3$. Also, we have reported the direct MCE results for the second-order material La$_{0.8}$Sr$_{0.2}$MnO$_3$.

\section{Instrumentation}

\begin{figure*}[ht]
    \centering
    \includegraphics[width=\linewidth]{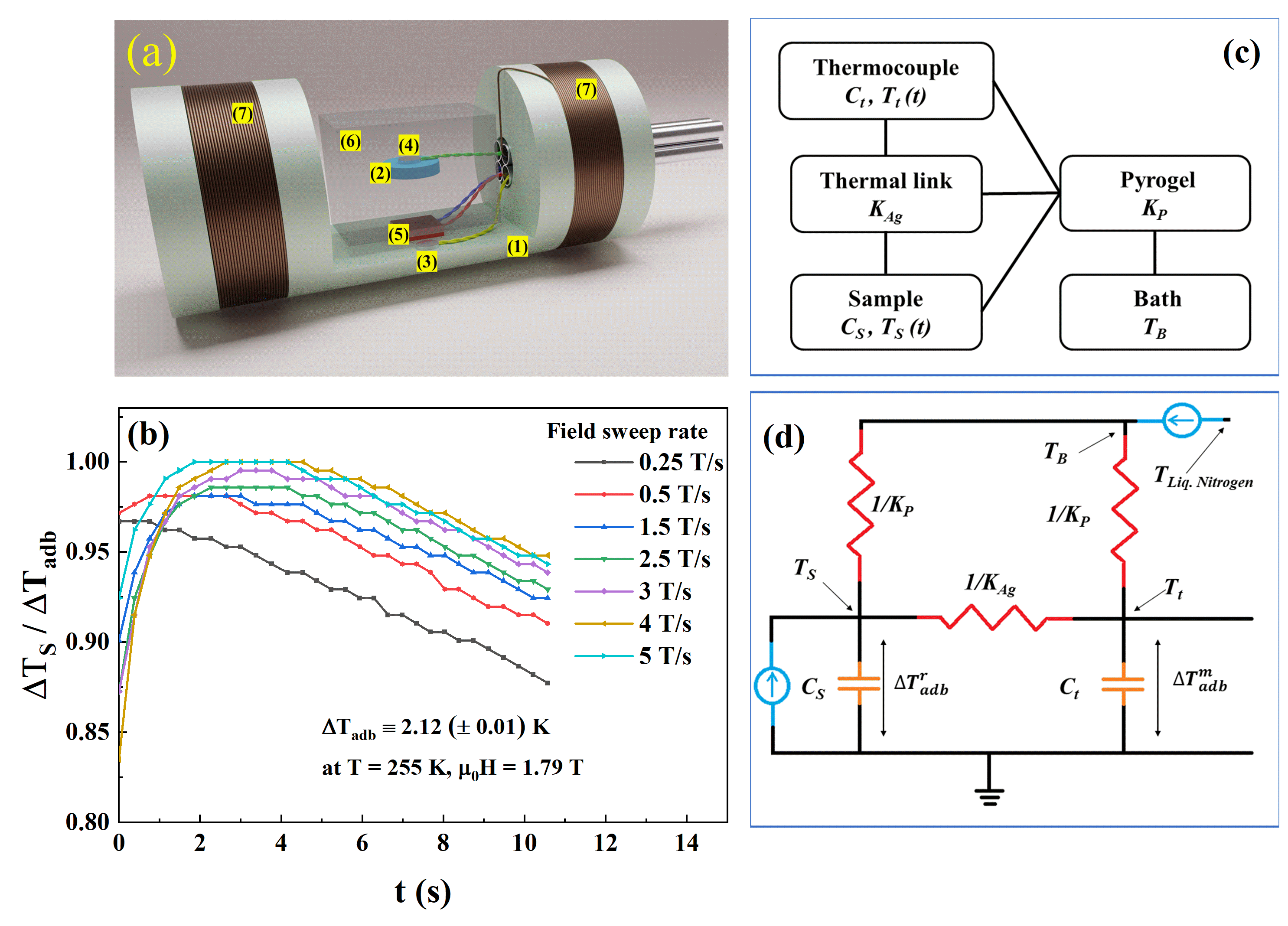}
    \caption{(a)Schematic of the sample rod used for $\Delta T_{ad}$ measurements; (1) brass sample holder, (2) sample, (3) thermocouple A  attached with the sample holder, (4) thermocouple B attached with the sample, (5) Hall sensor, (6) Pyrogel\textsuperscript{\textregistered} (7) Manganin\textsuperscript{\textregistered} heater. (b) Temporal variation of the sample temperature $\Delta T_{S}(t)/\Delta T_{adb}$ for different magnetic field sweep rates. (c) Block diagram of the heat transfer process in the system. (d) Equivalent electric circuit diagram of the system.}
    \label{Fig.schematic}
\end{figure*}

A $\Delta T_{adb}$ measurement process consists of three important steps; first adiabatic conditions should be established, second the magnetic field should be changed and measured, and third the temporal variation of the sample temperature should be monitored. These three steps have been incorporated in our setup. To create adiabatic conditions for the measurements, a high vacuum chamber ($\sim 10^{-6}$ hPa) is used. Along with the high vacuum chamber, the sample is wrapped with a layer ($\sim 2$~mm) of Pyrogel\textsuperscript{\textregistered} (cf. Fig. \ref{Fig.schematic}(a)) to reduce the heat transfer rate from the sample, which provide sufficient time for the measurement of $\Delta T_{adb}$  upon a change of the magnetic field. A schematic of the sample rod is shown in Fig. \ref{Fig.schematic}(a). The sample rod is placed inside the vacuum chamber, which is kept inside a liquid N$_2$ filled cryostat. Therefore, the system can operate at temperatures above the boiling point of liquid N$_2$.

The magnetic field variation was controlled by a Halbach-type permanent magnet array,\cite{tura2011permanent,coey2002permanent} which can produce a maximum magnetic field of $1.8$~T. To measure the magnetic field accurately at the position of the sample, a calibrated Hall-sensor is mounted on the sample rod, in a way that the  magnetic field is perpendicular to the Hall-sensor (cf. Fig. \ref{Fig.schematic}(a)). Although the heat radiation and convection are much reduced, the heat conduction from sample cannot be completely eliminated. This limits the overall measurement time. The measurement time is highly influenced by the magnetic field sweep rate, i.e. if the magnetic field sweep rate is low compared to the heat conduction rate from the sample, some amount of heat will be lost during the measurement. Khovaylo \textit{et al.} showed \cite{khovaylo2008adiabatic} that approximately a minimum magnetic field rate of $3$ T/s is required for a correct determination of $\Delta T_{adb}$. Using our system, the temporal variation of the sample temperature,  $\Delta T_{S}(t)$ have been measured for different magnetic field sweep rates for a first-order material, La$_{0.7}$Ca$_{0.3}$MnO$_3$. The measurements were  performed at a temperature near to the magnetic ordering temperature $T_C$ ($251$ K) of the material with the highest available magnetic field change ($1.79$ T). The recorded data are presented in Fig. \ref{Fig.schematic}(b). To describe the effect of magnetic field sweep rate, the temporal variation of the sample temperature has been normalized with the value of $\Delta T_{adb}$ ($2.12 (\pm 0.01)$ K) defined as the largest change of the sample temperature induced by a magnetic field change. From Fig. \ref{Fig.schematic}(b), it is clear that a minimum magnetic field sweep rate of $3$ T/s  is required to achieve a value of $\Delta T_{S}(t)/\Delta T_{adb}$ greater than $98\%$. Noticeably, a higher magnetic field sweep rate allows a longer time period for the $\Delta T_{adb}$ measurement. Therefore, all results presented in the following discussion have been obtained using a magnetic field sweep rate of $5$ T/s. This rate is high enough to neglect the heat dissipation during the magnetization or demagnetization process.

The temperature of the sample is monitored and controlled by a commercially available temperature controller (LakeShore 335). As temperature sensors, T-type thermocouples (accuracy of $\sim 0.01$ K) are used and as heat source a resistive Manganin\textsuperscript{\textregistered} heater providing a maximum  power of $50$ W is used. Both in thermocouples and in heater, twisted types of wires have been used in order to reduce any noise produced by the induced stray magnetic field in the wires\cite{gonschorek2010magnetic}. To understand the heat transfer process in the system, a block diagram and its equivalent electrical circuit are  presented in Figs. \ref{Fig.schematic}(c) and (d), respectively. The sample with heat capacity $C_S$ and temperature $T_S$ is attached to a thermocouple via a thermal link (Ag-paint with thermal conductivity $K_{Ag}$). Therefore, the temperature of the thermocouple $T_t$  differ slightly from the sample temperature $T_S$ and this difference depends upon the heat capacity of the thermocouple $C_t$. In the equivalent circuit, the inverse of the thermal conductivity represents a resistance, the heat capacitiy corresponds to a capacitance and the flow of heat is described by  an electrical current. Thus, the heat flow across the sample due to the change of magnetic field has been replaced by a current source. Moreover, there is heat flow from the sample to the liquid N$_2$ chamber through the Pyrogel\textsuperscript{\textregistered} (with thermal conductivity $K_P$), and this heat flow is being controlled by the resistive heater;  all these heat flows are collectively replaced by another current source in the equivalent electric circuit diagram. When the sample is subjected to a magnetic field change, either applied or removed, the temperature change of the sample is represented as $\Delta T_{adb}^{r}$, i.e. the real value of the adiabatic temperature change, while the measured temperature change across the thermocouple is represented as $\Delta T_{adb}^{m}$. These two quantities correspond to voltages across the capacitances $C_S$ and $C_t$ in the equivalent circuit and are related by the following equation,
\begin{equation}
    \Delta T_{adb}^{m} = \Delta T_{adb}^{r} \frac{C_S}{C_S + C_t} .
\end{equation}

Therefore, for an ideal measurement of $\Delta T_{adb}$, the value of $C_S$ should be much larger than the value of $C_t$. One way to satisfy this condition is to use larger mass of the sample compared to the mass of the thermocouple. Porcari \textit{et al.}\cite{porcari2013direct} has demonstrated experimentally the effect of sample mass on the $\Delta T_{adb}$ measurement, according to which around $50$~mg of sample is required for a measurement with better than $98\%$ accuracy. Apart from the sample mass, the thermal conductivity of the thermal link (Ag-paint) plays a crucial role in determining the time response of the $\Delta T_{adb}$ measurement. The time(t) response from the RC-circuit can be expressed as,
\begin{equation}
    \Delta T_{adb}^{m} = \Delta T_{adb}^{r} (1-e^{-t/\tau_t}),
\end{equation}

where $\tau_t$ is the time constant of the thermocouple.   $\tau_t$ determines the rate at which thermocouple temperature $T_t$ will reach the sample temperature $T_S$. From the equivalent RC-circuit, $\tau_t$ can be expressed as,

\begin{equation}
    \tau_t = \frac{C_t}{K_{Ag}}.
\end{equation}

Near room temperature, considering a T-type thermocouple and the thermal conductivity of silver paint, the value of $\tau_t$ is $<10^{-3}$~s. From Fig. \ref{Fig.schematic}(b) it is clear that the value of $\Delta T_{adb}$ is stable over a time period of a few seconds, which proves that our measurement process is reliable. 

Although Pyrogel\textsuperscript{\textregistered} is required to establish adiabatic conditions, it creates a time lag between the thermocouple attached to the sample and the heat source (resistive heater) which affects the process of temperature control. To overcome this problem, a second thermocouple has been introduced outside the Pyrogel\textsuperscript{\textregistered} attached to the brass sample holder. For further discussion, the thermocouple attached with the sample is named as thermocouple B, while the thermocouple attached to the sample holder will be referred to as thermocouple A. To  stabilize the sample temperature at any desired temperature, first the system temperature as measured by thermocouple A  is stabilized followed by a waiting time of a $\sim 10$ minutes until thermocouple B also shows the desired temperature and has been stable at this temperature for a few minutes.

\section{Measurement Process}

For the measurements in the direct measurement setup  two compounds have been selected, La$_{0.7}$Ca$_{0.3}$MnO$_3$ (LCMO), which shows first-order magnetic phase transition from the paramagnetic (PM) to ferromagnetic (FM) state at a temperature around $250$ K, and La$_{0.8}$Sr$_{0.2}$MnO$_3$ (LSMO), which shows a second-order PM to FM phase transition at a temperature around $325$ K. We have chosen two oxide materials instead of any intermetallic compound, as the oxide materials are advantageous owing to their chemical stability, ease to synthesis, resistance to corrosion etc. Also, the magnetocaloric intermetallic compounds are highly sensitive to the synthesis process and often incorporate more than one impurity phase. For instance, the giant magnetocaloric compound FeMnP$_{0.5}$Si$_{0.5}$ exhibits a $T_C$ of $\sim390$ K when synthesized by the drop synthesis method,\cite{hoglin2011crystal,ghorai2022site}  while  $T_C$ decreases to $\sim300$ K when synthesized using ball-milling\cite{cam2008structure}. Both materials studied in this work were synthesized using a modified sol-gel method described in details elsewhere\cite{ghorai2022effect}. Before describing the measurement results of these two compounds, a description of the $\Delta T_{adb}$ measurement process will be presented. 
\begin{figure}[ht]
    \centering
    \includegraphics[width=\linewidth]{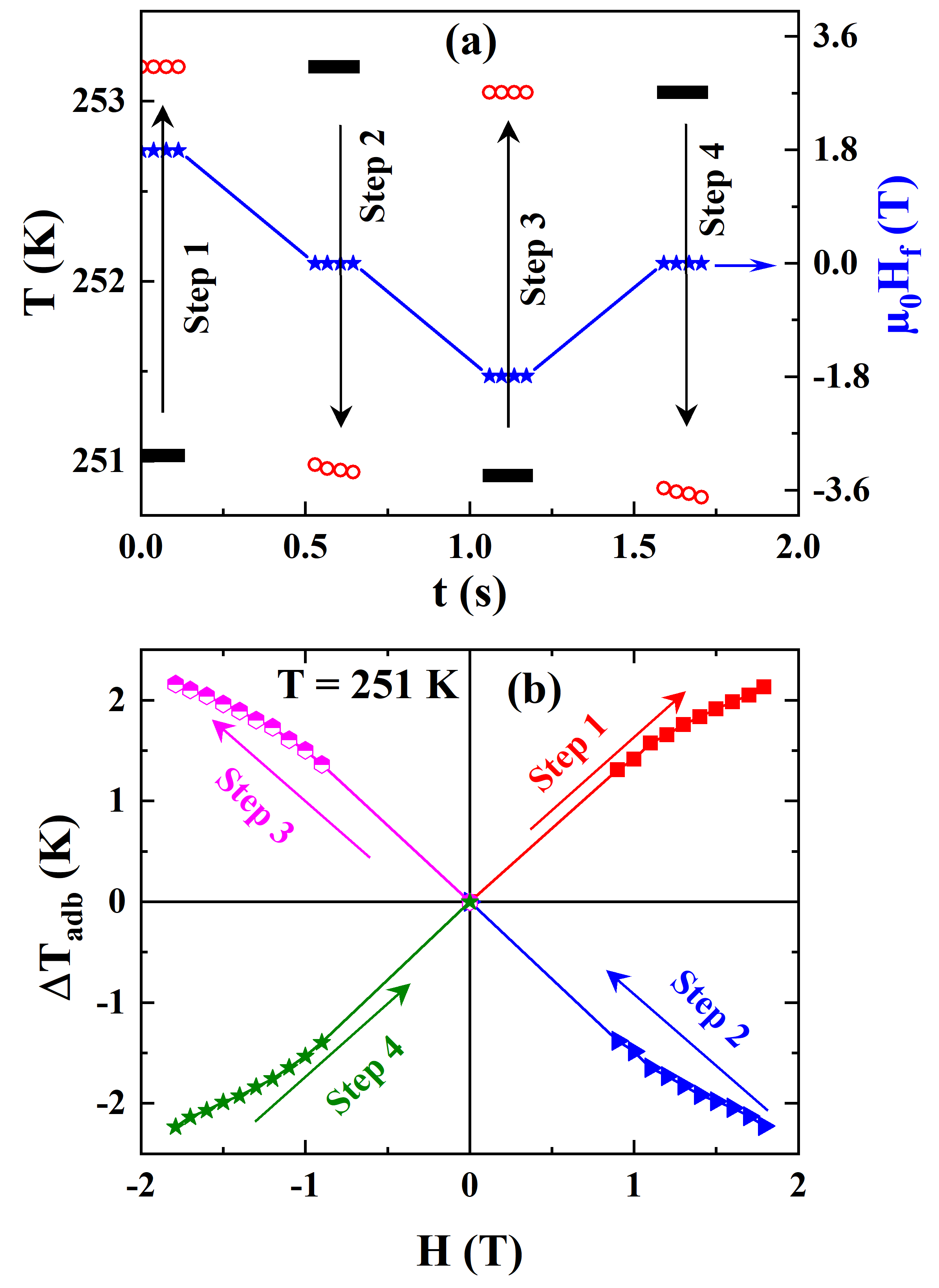}
    \caption{(a) A real time loop measurement protocol of magnetic field dependent temperature change for the LCMO compound. (b) Magnetic field dependent average values of $\Delta T_{adb}$ for different directions of the magnetic field change. Arrows indicate the magnetic field change direction.}
    \label{Fig.loop_measurement}
\end{figure}

Two types of measurement protocols have been used in this study, a loop measurement protocol and a phase reset protocol. The loop measurement protocol\cite{lyubina2010novel} corresponds to a fast measurement process of $\Delta T_{adb}$, which involves a continuous magnetization and demagnetization process while cooling or heating the compound. A schematic of the measurement protocol is presented in Fig. \ref{Fig.loop_measurement}(a) for the cooling cycle of the LCMO compound. Before starting the loop measurement, the sample temperature was stabilized at $T_1$ in zero applied magnetic field. At $T_1$, a positive magnetic field ($\mu_0H_f$) was applied raising the temperature of the sample to $T_1 + \Delta T_1$; this is indicated as step $1$ in Fig. \ref{Fig.loop_measurement}(a). In the second step, the magnetic field was changed to zero and consequently the sample temperature decreased from $T_1 + \Delta T_1$ to $T_1$. Steps $3$ and $4$ correspond to repetitions of steps $1$ and $2$, but this time a negative magnetic field with the same magnitude ($-\mu_0H_f$) was applied and subsequently removed. To minimize the statistical error in the estimation of $\Delta T_{adb}$, the temperature after every field change  was  measured at least four times over a time span of $\sim 1~s$ and the average value was considered as the final value of $\Delta T_{adb}$. Figure \ref{Fig.loop_measurement}(b) shows the final values of $\Delta T_{adb}$ at $251$ K for the LCMO compound with cyclic field changes from $0$ T to $1.79$ T (step $1$), $1.79$ T to $0$ T (step $2$), $0$ T to $-1.79$ T (step 3) and finally $-1.79$ T to $0$ T. 
The second measurement protocol, the phase reset protocol, is a modified version of the loop measurement protocol. In the phase reset protocol, before stabilizing the sample at the temperature $T_1$, the sample is heated (cooled) to a temperature in the PM (FM) state for cooling (heating) cycle measurements. Carron \textit{et al.}\cite{caron2009determination} have shown that the field and temperature hysteresis of a first-order compound plays an important role and without adopting  the phase reset protocol, the calculation of the isothermal entropy change using Maxwell's relation often generates a large error. The error involved in the direct measurement of $\Delta T_{adb}$ is related to the accuracy of the absolute temperature measured by the thermocouple,  the magnetic field dependence of the thermocouple and possible heat dissipation during measurement due any non-adiabatic condition. However, as the thermocouple mass is negligible compared with the mass of the sample and a T-type thermocouple with negligible magnetic field dependence is used, the error is significantly reduced. The  negligible effect of magnetic field on the thermocouple has been verified by performing temperature dependent loop measurements without a sample. Moreover, the measurement of $\Delta T_{adb}$ involves a time span (after application or removal of the magnetic field) where the temperature change is constant (cf. Fig. \ref{Fig.schematic}(b)), i.e. no heat loss is observed, which has been confirmed by repeating the measurement at least four times. Therefore, the only error involved in this measurement is due to the resolution of the temperature sensor, which is in the order of $\sim 0.014$ K near room temperature.

\section{Data analysis}

\begin{figure}[ht]
    \centering
    \includegraphics[width=\linewidth]{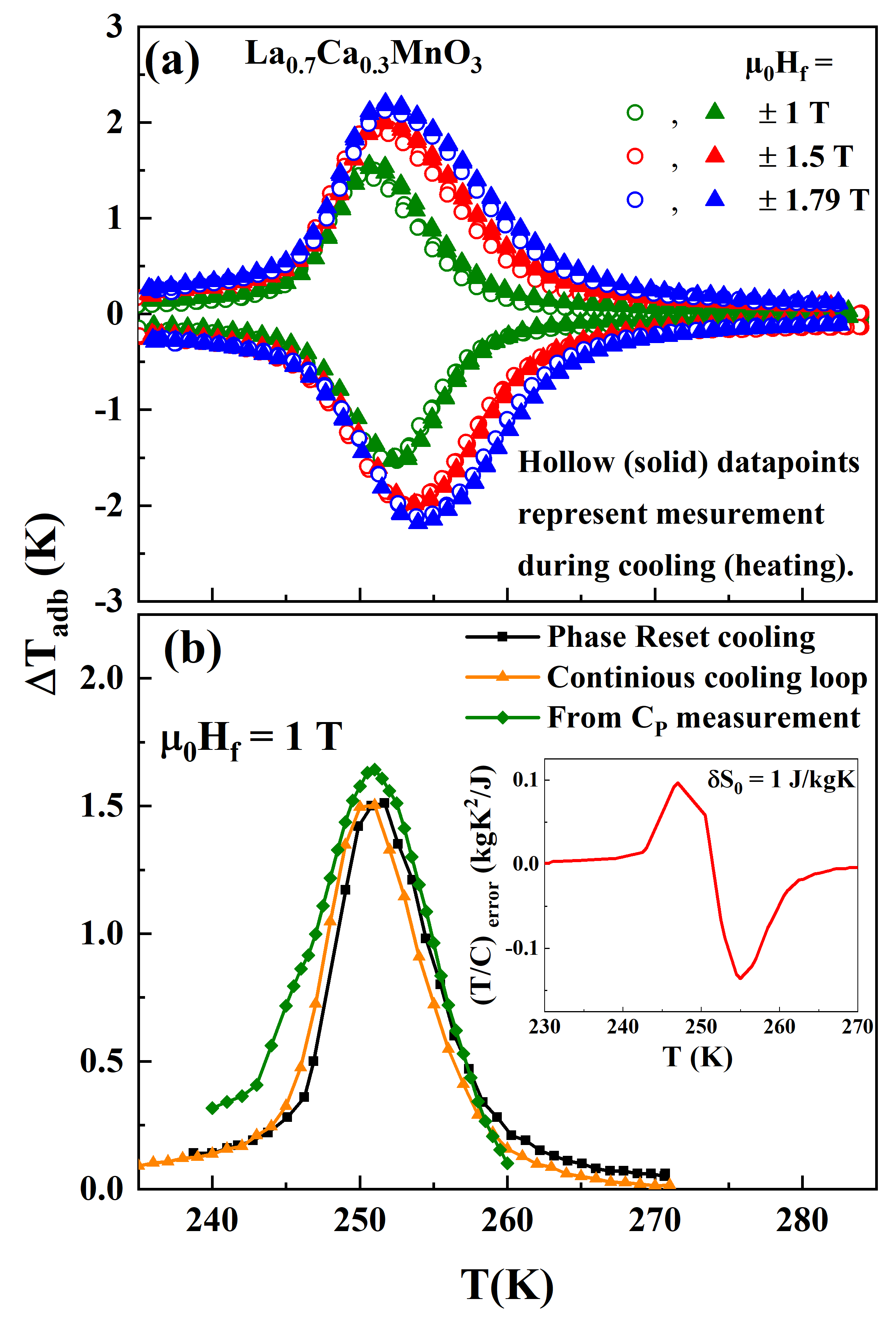}
    \caption{(a) Temperature dependent variation of $\Delta T_{ad}$ for the LCMO compound for cooling and heating cycles. The different curves correspond to different magnetic fields $\mu_0H_f$. The positive (negative) values of $\Delta T_{ad}$ correspond to magnetization (demagnetization) of the compound. (b) Comparison of  $\Delta T_{ad}$ measured following continuous cooling and phase reset protocols along with the calculated values from the indirect measurement. The inset shows the temperature dependence of the error term $(T/C)_{error}$.}
    \label{Fig.LCMO}
\end{figure}

Figure \ref{Fig.LCMO}(a) shows an example for the LCMO compound of the temperature dependent loop measurement protocol. LCMO exhibits a first-order magnetic transition, i.e. there is a discontinuity in the first derivative of the Gibbs free energy. The first-order nature of the magnetic transition for this compound has been verified using the Arrott plot analysis (the result of the analysis is presented in SI).\cite{arrott1967approximate,ghorai2020field} During the magnetization process with both positive and negative magnetic fields the temperature of the sample increases, while it decreases during the demagnetization process. The magnetic field dependent shift of the peak maxima during magnetization and demagnetization corresponds to the field hysteresis of the compound near to $T_C$. The origin of the field hysteresis is the magnetic field dependent metamagnetic behaviour of the LCMO compound.\cite{ulyanov2006giant} Apart from the field hysteresis, there is a small shift of the $\Delta T_{adb}(T)$ curves comparing results collected during the heating and cooling cycles, i.e. the compound also exhibits a temperature hysteresis. From the temperature dependent magnetization data (results presented in SI) it is confirmed that there is an approximately $1$ K temperature difference in the value of $T_C$ comparing heating and  cooling cycles, which causes the temperature hysteresis in the $\Delta T_{adb}$ measurement. This temperature hysteresis is an unavoidable consequence of the first-order magnetic transition. Another consequence of the first-order magnetic transition is the shift of $T_C$ towards higher temperature with increasing magnetic field. As a result the $\Delta T_{adb}$ curves shift towards higher temperature with increasing field, confirming the expected behaviour for a first-order magnetic phase transition.\cite{lyubina2017magnetocaloric,titov2012hysteresis}

The field dependence of the transition temperature $T_C$ has direct consequences on the measurement protocols. In Fig.\ref{Fig.LCMO}(b), a comparison between the data collected following the continuous cooling and the phase reset protocols is presented. The only noticeable change is in the $\Delta T_{adb}$ peak temperature, not in the peak amplitude. Caron \textit{et al.}\cite{caron2009determination} have shown that for a first-order compound, the calculated value of the isothermal entropy change following the phase reset protocol is almost half of the value calculated using the continuous cooling protocol. As a reason for this large  difference of isothermal entropy change, they pointed out that the basic assumption in Maxwell's relation, 
\begin{equation}
    \pdv{S}{H}_T = \pdv{M}{T}_H,
\end{equation}
is not valid for first-order magnetic transitions. Fortunately, in the direct measurement there is no such assumption made and as a result both the continuous cooling and the phase reset protocols indicate same value of $\Delta T_{adb}$. 

To understand the peak temperature shift between the two measurement protocols, consider  the continuous cooling method. The sample is cooled from the PM state to a temperature $T_1$ in the magnetic transition region ($T_1 > T_C$), followed by the application of a magnetic field. As $T_1 > T_C$, the magnetic moments in the compound will be partially aligned by the magnetic field. However, when the field is removed, the magnetization of the compound will not be zero. Now, if a second temperature $T_2$ ($T_1 > T_2 > T_C$) is set following  the continuous cooling protocol and a magnetic field of the same magnitude is applied the magnetization at temperature $T_2$ will be affected by the remanent magnetization from temperature $T_1$. The  effect of the remanent magnetization will always be present unless the compound is heated to its PM state. Thus, in the continuous cooling process the $\Delta T_{adb}(T)$ curve will be influenced by the temperature dependent remanent magnetization  contribution. On the other hand, the phase reset cooling process only involves the magnetic field dependent metamagnetic phase transition. This explains the peak temperature shift between the two measurement protocols. It can be concluded that the phase reset protocol should be adopted for a first-order compound. However, for a real magnetic refrigeration cycle, for example the AMR (Active Magnetic Regenerative) refrigeration cycle ,\cite{el2019magnetocaloric,aprea2016rotary} the involved cooling process is the continuous cooling process. Therefore, while measuring $\Delta T_{adb}$, data from both protocols should be presented for basic understanding of the magnetic phase transition as well as for the application of the material in refrigeration cycles.

To compare the direct and the indirect measurement data of the MCE, temperature dependent specific heat measurements at two magnetic fields, $0$ T and $1$ T, have been performed. In the indirect measurement method, the total entropy ($S_H(T)$) is calculated from the specific heat data by numerical integration of the thermodynamic relation, 

\begin{equation}
    S_H(T) = \int_{0}^{T} \frac{C_H(T)}{T} \,dT + S(0),
\end{equation}

where $S(0)$ is the magnetic field independent zero-temperature entropy. The $C_H(T)$ data have been collected using the heat capacity option of the Quantum Design PPMS\cite{rosen2020standard}. The calculation is based on  the assumption $C_H(0)=0$, which together with the unknown zero-temperature entropy will introduce a temperature and magnetic field independent constant error $\delta S_0$. The adiabatic temperature change  for the field change $0 \rightarrow H_f$ can then be  calculated using the relation,\cite{pecharsky1999magnetocaloric},

\begin{equation}
\label{Eq.5}
    \Delta T_{adb}(T,H_f) = [T(S_{H_f}(T)+\delta S_0) - T(S_{0}(T)+\delta S_0)]_S.
\end{equation}

 Taylor expanding the functions $T(S_{H}(T)+\delta S_0)$ on the right-hand side of Eq.\ref{Eq.5}, keeping only the first term in the expansion and using the relation $[dT(S)/dS]_H = [T/C(T)]_H$, Eq. \ref{Eq.5} can be expressed as,\cite{pecharsky1999magnetocaloric}

\begin{equation}
\label{eq.6}
\begin{split}
    \Delta T_{adb}(T,H_f) = \left[T(S_{H_f}) - T(S_{0})\right]_{S}\\+\delta S_0\left[\frac{T}{C_{H_f}(T)}-\frac{T}{C_0(T)}\right]_{S}.
\end{split}    
\end{equation}

The second term on the right hand side of Eq.\ref{eq.6} indicates the possible error involved in the calculation of $\Delta T_{adb}$. Although  the value of $\delta S_0$ is constant, the term \(\left[\frac{T}{C_{H_f}(T)}-\frac{T}{C_0(T)} \right]_{S}\) is temperature and field dependent; this term will in the following be referred to as $(T/C)_{err}$. In Fig. \ref{Fig.LCMO}(b), the calculated values of $\Delta T_{adb}$, ignoring the error contribution from Eq.\ref{eq.6}, is shown. In the  inset of Fig. (\ref{Fig.LCMO}(b)), the temperature dependence  of $(T/C)_{err}$ is shown using $\delta S_0 = 1$ J/kgK, clearly showing that the error involved in Eq.\ref{eq.6} will have a positive (negative) contribution on the calculated value of $\Delta T_{adb}$ below (above) $T_C$. However, Pecharsky and Gschneidner\cite{pecharsky1999magnetocaloric} have pointed out that this error term can be ignored as the value of $\delta S_0$ is considerable small. Considering a random error for the indirect measurement yields a large error for $\Delta T_{adb}$ (see SI) of the order of $1$ K, while the error involved in direct measurement is of the order of $0.01$ K. This reinforces the notion that direct measurements of the MCE are needed  for the correct determination of $\Delta T_{adb}$. Apart from the random error, there is an experimental limitation with the grease used to mount the sample in specific heat measurements. Two greases are typically available for the specific heat measurement, the N-grease that can be used below room temperature and the H-grease that should be used above room temperature. Therefore, materials with a $T_C$ above room temperature require  separate measurements with the two greases  and each measurement consists of one addenda (background) and one sample measurement. A total of four specific heat measurements is therefore needed to cover the desired temperature range. Noticeably, for the determination of $\Delta T_{adb}$, the specific heat should be measured for two magnetic fields. This makes a total of eight separate measurements and all measurements  should be performed under the same  conditions. This makes the indirect measurement challenging for materials with a $T_C$ above room temperature. 

\begin{figure}[ht]
    \centering
    \includegraphics[width=\linewidth]{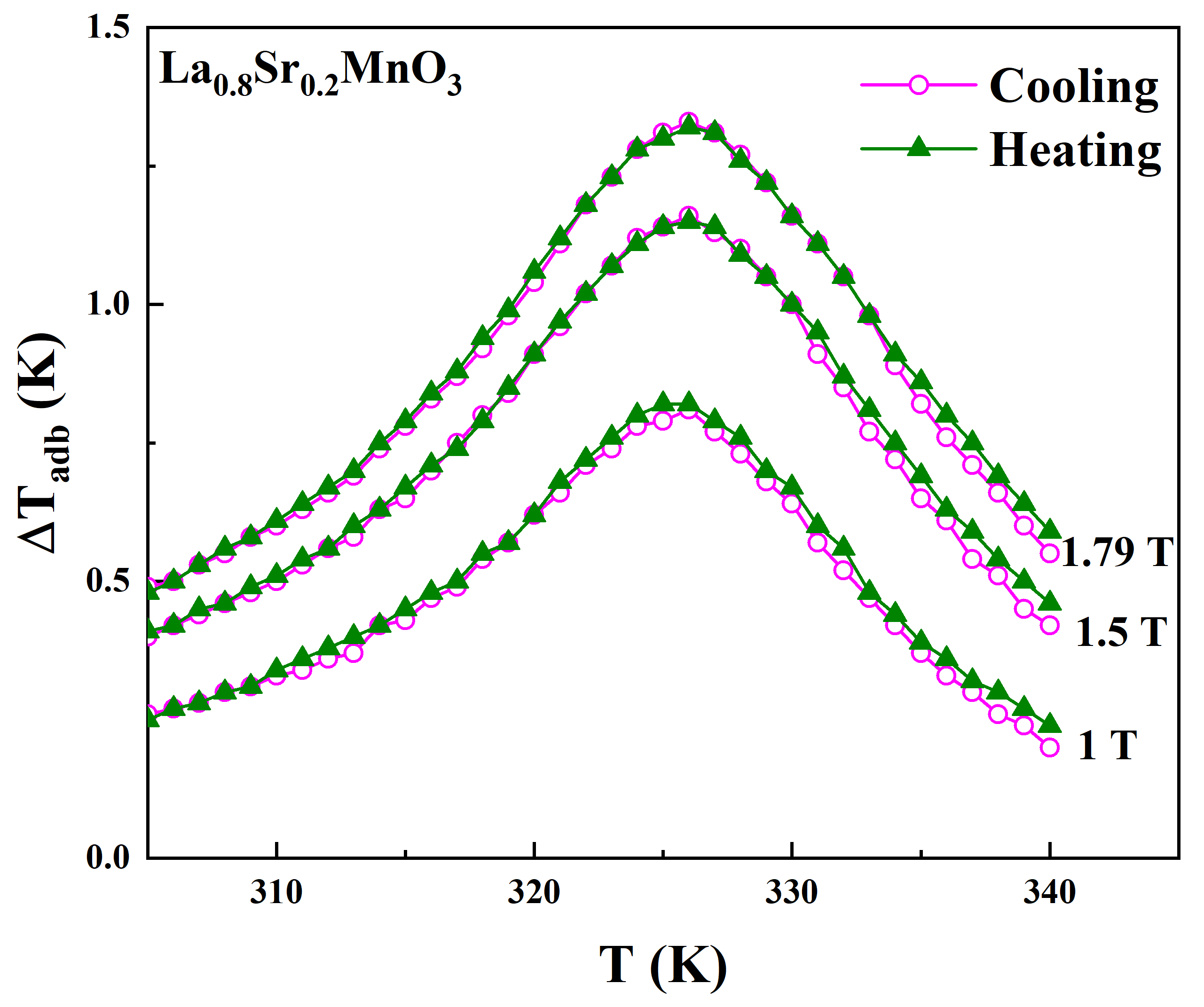}
    \caption{Temperature dependent variation of $\Delta T_{ad}$ for the second-order material LSMO for cooling and heating cycles.}
    \label{Fig.2nd_order}
\end{figure}

So far, results from direct measurements of the MCE for a first-order compound have been discussed. As a comparison, direct measurements of the MCE have also been performed  for the second-order material LSMO. The basic magnetic properties of this compound have been reported elsewhere\cite{ghorai2022effect}. Figure \ref{Fig.2nd_order} shows the temperature dependent variation of the $\Delta T_{adb}$  for this compound measured in the loop protocol. As expected, there is no hysteresis in heating and cooling cycles for a second-order material.

\section{\label{sec:conclusion}Conclusions}

A simple setup for the measurement of  $\Delta T _{adb}$ has been demonstrated here. It was observed that a rate of magnetic field change of $\sim3$ T/s is required for a measurement of $\Delta T_{adb}$ with $2 \%$ relative error. The $\Delta T _{adb}$ results for the first-order compound La$_{0.7}$Ca$_{0.3}$MnO$_3$ exhibit temperature and field hysteresis owing to the temperature dependent irreversibility of the magnetic phase transition and the metamagnetic behaviour of the compound. The result for  $\Delta T _{adb}(H,T)$ from direct measurements has been compared with the indirect result obtained from the specific heat results. Although the results from two measurements are in good agreement near $T_C$, the large relative error in the indirect measurement indicates that the direct measurement is superior to the indirect measurement. Apart from the data quality in the indirect measurement, it is also a relatively time consuming measurement. To obtain $\Delta T_{adb}$ for one magnetic field change a measurement time of around $2$ days or more is required. As a comparison, the direct measurement setup demonstrated here can measure the same within a few hours. Moreover, $\Delta T_{adb}$ of a second-order compound has also been presented and the temperature dependent reversibility of the second-order compound has been verified.

\begin{acknowledgments}
The authors thank the Swedish Foundation for Strategic Research (SSF), project ``Magnetic materials for green energy technology” (contract EM$-16-0039$) for financial support. The authors would also thank the project students Ronja Olsmats Baumeister, Oliver Djurle, Ella Kurland, Sebastian Åberg and Axel Gudmundsson. Finally, the authors would like to give thanks to Dan Bergman and Niklas Johansson from the Ångstr{\"o}m workshop for practical issues as well as technical drawings.
\end{acknowledgments}

\bibliography{directMCE.bib}% Produces the bibliography via BibTeX.
\bibliographystyle{prb-titles.bst}
\begin{widetext}
\clearpage
\appendix

\section{Supplementary Information}

The basic magnetic properties of the LCMO compound are presented in Fig. (\ref{mag_supp}). From the temperature dependent magnetization, a small temperature hysteresis ($\sim 1$ K) is observed. The negative slopes in the Arrott plot analysis confirm the first-order nature of the LCMO compound. The temperature dependence of the total entropy calculated using results from the specific heat measurements are shown in Fig.\ref{mag_supp}(d) for two different magnetic fields. Within the margin of error, the change in entropy with magnetic field is only observed in the vicinity of $T_C$, which explains why both the isothermal entropy change and the  adiabatic temperature change are zero well below and well above the magnetic phase transition temperaure.

\begin{figure}[ht]
    \centering
    \includegraphics[width=\linewidth]{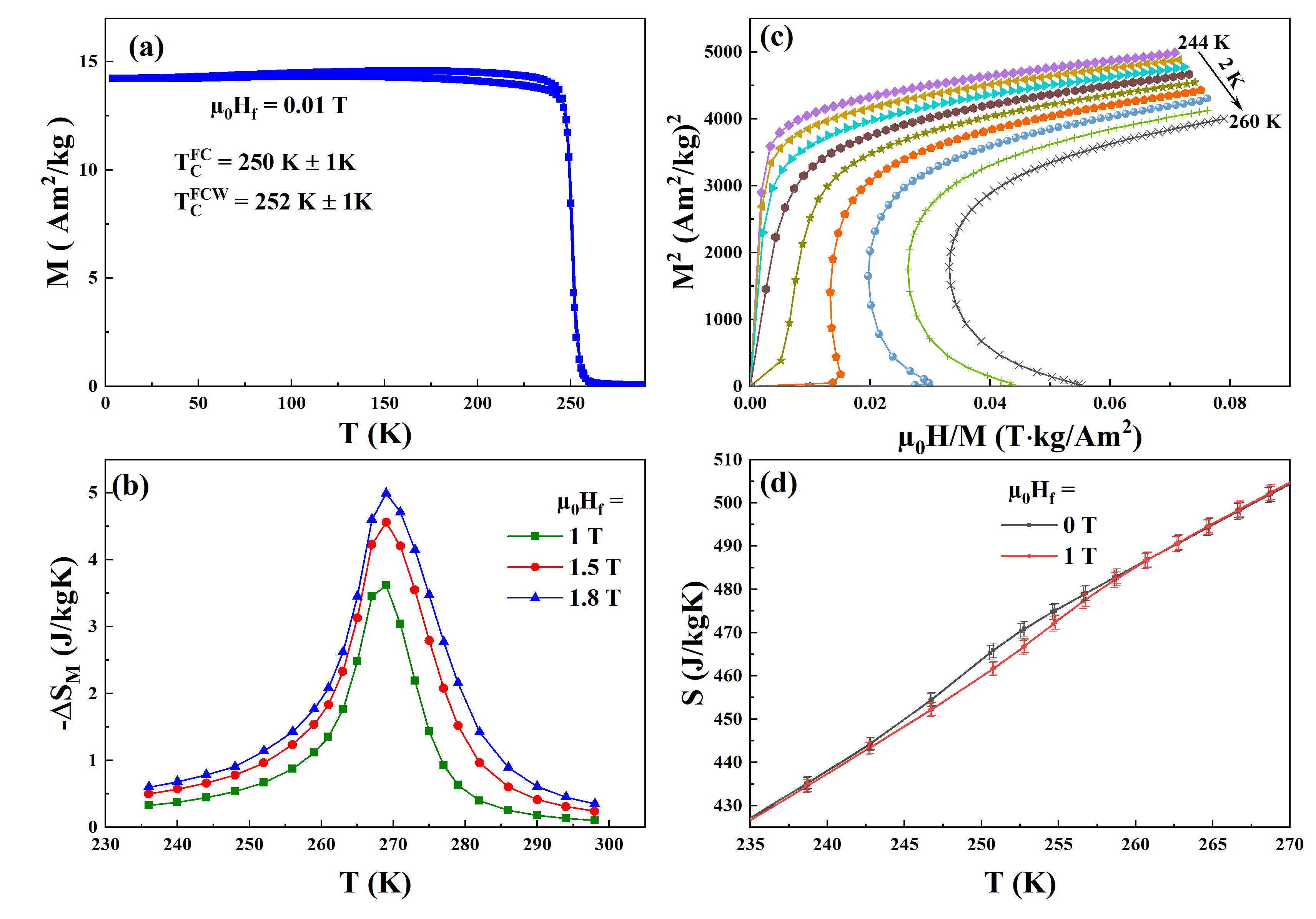}
    \caption{(a) Temperature dependent variation of the magnetization measured in field cooled cooling and field cooled warming modes, (b) temperature and field dependent isothermal entropy change, (c) Arrott plots  and (d) temperature dependence of the total entropy at two different magnetic fields.}
    \label{mag_supp}
\end{figure}

In Fig. \ref{random_error}, the temperature dependent values of $\Delta T_{adb}$ for the LCMO compound along with the calculated random error in the indirect measurement are shown. Following Pecharsky \textit{et al.}\cite{pecharsky1999magnetocaloric}, the random error  $\sigma\Delta T_{adb}$ can be defined as,

\begin{equation}
    \sigma\Delta T_{adb}(T,H_f) = \left[\sigma S_{H_f}(T)\frac{T}{C_{H_f}(T)}+\sigma S_{0}(T)\frac{T}{C_{0}(T)}\right]_{S}, 
\end{equation}

where $\sigma S_{0}(T)$ and $\sigma S_{H_f}(T)$ are the random errors of the total entropy at zero magnetic field and at the magnetic field $H_f$.

\begin{figure}[ht]
    \centering
    \includegraphics[width=0.6\linewidth]
    {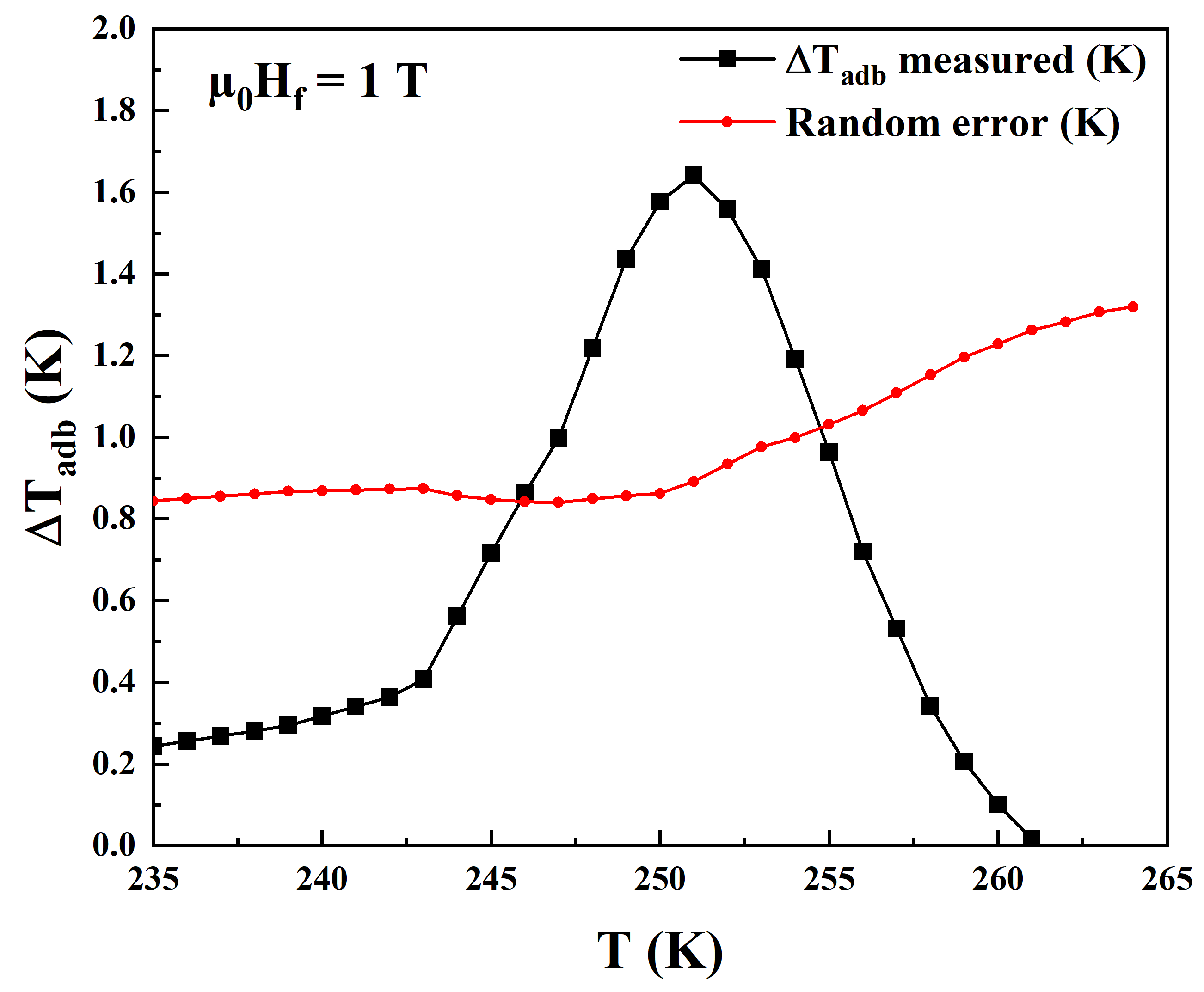}
    \caption{(a) Temperature dependent values of $\Delta T_{adb}$ derived from the indirect measurements along with the calculated random error for the LCMO compound.}
    \label{random_error}
\end{figure}
\end{widetext}

\end{document}